\documentclass[twocolumn]{aastex62}
\usepackage{graphicx}
\usepackage{caption2}

\begin{document}

\title{The role of multiple giant impacts in the formation of the Earth-Moon system}

\correspondingauthor{Robert Citron}
\email{ricitron@berkeley.edu}

\author{Robert I. Citron}
\affiliation{Department of Earth and Planetary Science, University of California, Berkeley, CA, USA}

\author{Hagai B. Perets}
\affiliation{Department of Astrophysics, Israel Institute of Technology, Haifa, Israel}

\author{Oded Aharonson}
\affiliation{Department of Earth and Planetary Science, Weizmann Institute of Science, Rehovot, Israel}

\vspace{0.25in}

\begin{abstract}
The Earth-Moon system is suggested to have formed through a single giant collision, in which the Moon accreted from the impact-generated debris disk. However, such giant impacts are rare, and during its evolution the Earth experienced many more smaller impacts, producing smaller satellites that potentially coevolved. In the multiple-impact hypothesis of lunar formation, the current Moon was produced from the mergers of several smaller satellites (moonlets), each formed from debris disks produced by successive large impacts. In the Myrs between impacts, a pre-existing moonlet tidally evolves outward until a subsequent impact forms a new moonlet, at which point both moonlets will tidally evolve until a merger or system disruption. In this work, we examine the likelihood that pre-existing moonlets survive subsequent impact events, and explore the dynamics of Earth-moonlet systems that contain two moonlets generated Myrs apart. We demonstrate that pre-existing moonlets can tidally migrate outward, remain stable during subsequent impacts, and later merge with newly created moonlets (or re-collide with the Earth). Formation of the Moon from the mergers of several moonlets could therefore be a natural byproduct of the Earth's growth through multiple impacts. More generally, we examine the likelihood and consequences of Earth having prior moons, and find that the stability of moonlets against disruption by subsequent impacts implies that several large impacts could post-date Moon formation.
\end{abstract}


\section{Introduction}
The single giant impact hypothesis is the most prevalent theory of Moon formation because it explains the angular momentum of the Earth-Moon system and the Moon's depletion in iron and volatile elements \citep{Canup2004a}. Any impact hypothesis must also explain why the Earth and Moon have similar oxygen, tungsten, and titanium isotope ratios \citep{Wiechert2001,Zhang2012,Herwartz2014,Kruijer2015}, which would normally vary among planetary embryos \citep{Kaib2015a,Mastrobuono-battisti2015,Mastrobuono-Battisti2017}. Impact models can account for isotope similarities either via a gas-rich protolunar disk that allows the proto-Moon and proto-Earth to equilibrate \citep{Pahlevan2007, Salmon2012, Lock2017}, or via impact dynamics if the impactor was compositionally similar \citep{Mastrobuono-battisti2015,Mastrobuono-Battisti2017} or had higher angular momentum than the present state. High initial angular momentum could be subsequently dissipated via the evection resonance \citep{Cuk2012}, limit cycles \citep{Wisdom2015}, material ejection \citep{Reufer2012}, or a high-obliquity Earth \citep{Cuk2016}. 

\begin{figure*}
\centerline{\includegraphics[trim={0 1cm 0 1.5cm},clip,scale=0.9]{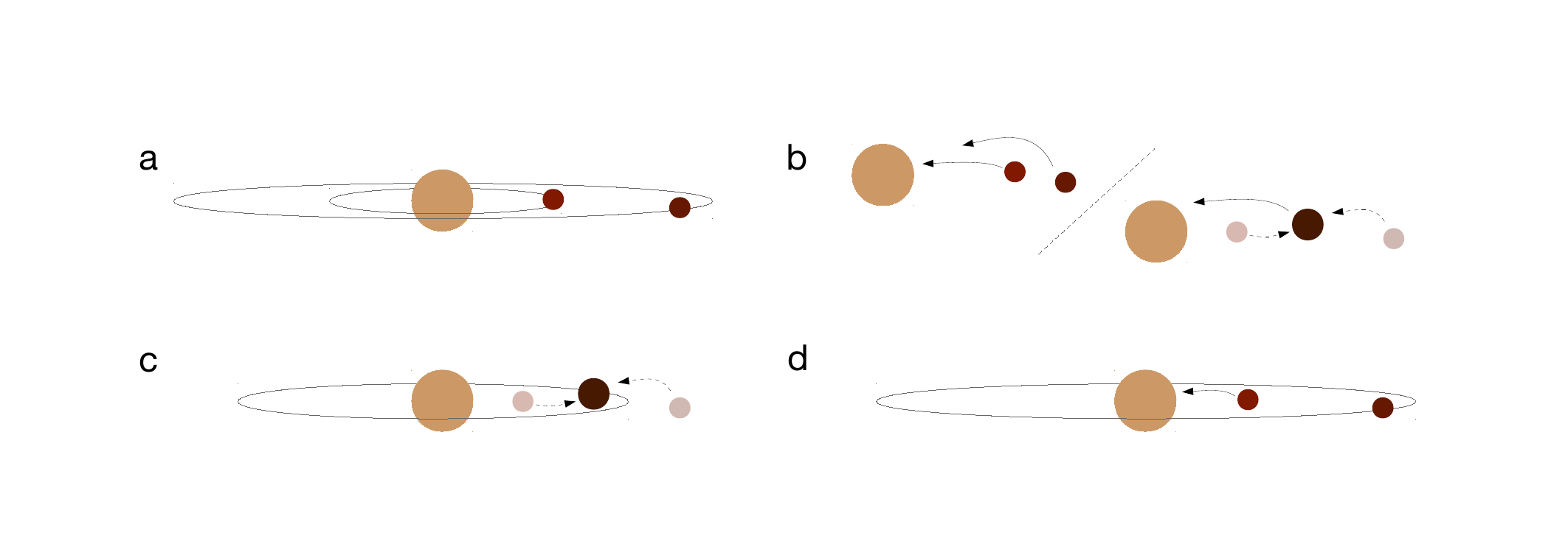}}
\caption{Potential outcomes from our simulations. The system begins with (a) an outer moonlet from a previous impact/merger that has migrated to a distant orbit, and an inner moonlet produced by a recent impact that begins at the Roche limit. The system evolves until either (b) both moonlets infall individually or after a merger, impacting the proto-Earth (in rare cases one moonlet can be ejected (not shown)), (c) the moonlets merge into a larger moonlet that remains stable and continues to tidally evolve outward, or (d) the inner moonlet infalls while the outer moonlet remains stable (in rare cases the outer moonlet infalls and the inner moonlet remains stable). Note that the starting inclinations need not be co-planar (as assumed in (a)), and that the final inclination of the moonlets could change. To build the Moon from multiple moonlet-generating impacts, the probability of mergers and outer moonlet survival (where the outer moonlet could be a product of a previous merger) must be high. } 
\label{fig:graphic}
\end{figure*}

However, the collisions necessary to form the Moon from a single impact are rare, appearing in only 2$-$8\% of $N$-body simulations from \citet{Brasser2013} and \citet{Elser2011}. This is partly due to the unlikelihood of forming an Earth-Moon system with a small inclination relative to the ecliptic when the distribution of impacts is isotropic \citep{Agnor1999,Chambers2001,Kokubo2007,Kokubo2010}, a constraint that may be relaxed in the context of a high-obliquity early Earth \citep{Cuk2016}. Late collisions of similarly sized bodies ({\it e.g.}, \citet{Canup:2012}) are also unlikely \citep{Jacobson2014a}, and the probability of a compositionally similar impactor is only $\sim$ 10\% \citep{Mastrobuono-Battisti2017}. Although standard and hit-and-run collisions ({\it e.g.}, \citet{Cuk2012} and \citet{Reufer2012}) are more feasible \citep{Jacobson2014a}, it is unclear how often they are the last impact, and impacts of less mass and higher impact velocity are more frequent \citep{Raymond2009a}. Furthermore, high angular momentum scenarios \citep{Cuk2012} require an initially fast-spinning Earth, a condition that is disfavored by angular momentum drain due to previous impacts \citep{Rufu2017}.

Such challenges have led some to propose a multiple-impact model of Moon formation \citep{Ringwood1989,Rufu2017}, in which the Moon forms as a natural consequence of the many large impacts the proto-Earth experienced during planetary formation. Each impact can produce a debris disk that spawns a small moonlet, which tidally evolves outward relatively quickly before subsequent collisions. If a new moonlet is formed by a subsequent collision, which is feasible because the 1-100 yr timescale of satellite formation from a debris disk \citep{Kokubo:2000,Salmon2012} is much shorter than the millions of years that pass between embryo-embryo collisions \citep{Raymond2009a,Morishima2010}, the result is a two-moonlet system with an older outer moonlet and a new inner moonlet. The two-moonlet system would dynamically evolve, potentially resulting in a merger. A sequence of new impacts and subsequent mergers could eventually build the single final Moon \citep{Rufu2017}. The similarity in isotope ratios between the Earth and the Moon develops over time as the Moon is built from moonlet mergers, and because the final angular momentum of the Earth-Moon system is the cumulative result of multiple impacts, smaller and faster impacts are allowed, which would eject more proto-Earth material into orbit. 

\citet{Rufu2017} showed that a range of sub-lunar mass debris disks can be produced by typical collisions during planetary accretion (0.01$-$0.1 Earth masses ($M_{e}$)), producing satellites 0.1 to 0.5 lunar masses ($M_l$). However, in order to explain the similarity in isotope ratios, \citet{Rufu2017} found that $\sim$ 20 mergers between impact generated satellites were required. Such a high number of impacts is possible according to $N$-body simulations \citep{Raymond2009a,Morishima2010}, where the average interval between large collisions (0.01$-$0.1$M_{e}$) is $\sim$ 6 Myr. Slightly less than 20 impacts is generally consistent with the formation of the Moon, estimated to be $\sim 95 \pm 32$ Myr after condensation based on compositional constraints \citep{Jacobson2014b}. However, forming the Moon from such a high number of impacts requires that the satellites produced from mergers of early satellites remain stable during subsequent giant impacts and dynamical interactions with newly generated satellites.

We further investigate the multiple impact hypothesis by examining the dynamics of multiple moonlet systems (Fig. \ref{fig:graphic}). While the evolution of multiple-moon systems has been studied before \citep{Canup1999,Jutzi2011}, such studies focused on a single impact genesis. We examine systems with an outer moonlet that migrated to its position during the interval between impacts, and a newly-generated inner moonlet that starts just outside the Roche limit. We first determine the probability that moonlet-generating impacts would disrupt the orbit of any pre-existing moonlet. Then, using $N$-body simulations with direct tidal forces/torques, we access the probability that two moonlets generated Myrs apart eventually merge.

\section{Pre-impact interactions}

\subsection{Methods}
To ensure no disruption of the proto-Earth--moonlet system occurs prior to the impact, we examined the pre-impact dynamics of single--binary collisions. Following \citet{Rufu2017}, we consider impactors of mass (0.01$-$0.1$M_{e}$), producing satellites 0.1$-$0.5$M_l$. Between consecutive impacts, a pre-existing moonlet would tidally migrate in the $\sim$ 6 Myr between impact events. Using a standard model for satellite tidal evolution \citep{Murray2000}, with tidal dissipation function $Q=34$ and degree-2 potential Love number $k_2 = 0.299$, a 0.1, 0.3, and 0.5 $M_l$ moonlet would migrate to $\sim$ 15, 18, and 20 $R_e$, respectively, in 6 Myr.

Using the direct $N$-body integrator Fewbody \citep{Fregeau2004}, we simulated single--binary encounters with a moonlet of mass 0.3 $M_l$ orbiting an Earth-mass body at 18 $R_e$. The impactor was given a mass of 0.01, 0.025, 0.05, or 0.1 $M_e$ and a velocity of 1, 1.4, 2, and 4 times the escape velocity (the range of velocities and masses considered by \citet{Rufu2017}). For each set of impact parameters, 1000 encounters were simulated with the Fewbody code, with random impact angles and initial orbits. The impact parameter $b$ was set to collisional trajectories of $b$ = 0 or 1 $R_e$, and non-collisional trajectories of $b$ = 0.5, 1, and 10 times the separation of the proto-Earth-moonlet binary (18 $R_e$), to test for disruptions from close encounters. The output from Fewbody was used to estimate the change in the orbital parameters of the pre-existing moonlet from the collisional or close encounter trajectory of the impactor. We repeated the calculations for a 0.1 and 0.5 $M_l$ moonlet, to examine the range of moonlets found to form in \citet{Rufu2017}.

\subsection{Results}
Our simulations show that collisions of an impactor with a proto-Earth--moonlet binary generally do not affect the orbit of a pre-existing moonlet. The results for a system with moonlet mass 0.3 $M_l$ and impactor with velocity $v_{imp}=1.4 v_{esc}$ are shown in Table \ref{tbl:fewbody}. For impactors with mass $M_i \le 0.05 M_e$, almost all collisional trajectories ($b$ = 0 or 1 $R_e$) resulted in the impactor colliding with the proto-Earth and the moonlet remaining in orbit. The orbital separation of the proto-Earth--moonlet pair slightly increased for $M_i = 0.05 M_e$. For a more massive impactor, $M_i$ = 0.1 $M_e$, collisional trajectories resulted in an original moonlet survival probability of 87--89\%, with the orbital separation of the binary increasing by a factor of $\sim$ 3. Following collisions, the average change in inclination of the pre-existing moonlet was $< 5^{\circ}$ for $M_i \le 0.05 M_e$ and $\sim 10^{\circ}$ for $M_i = 0.1 M_e$. When the moonlet did not maintain an orbit, it was either ejected from the system or also collided with the proto-Earth (a triple merger). For non-collisional trajectories, except in rare cases, the impactor passed through the system and the original binary was preserved. 

\begin{table*}
\centering
\begin{tabular*}{\textwidth}{@{\extracolsep{\fill}}c@{\extracolsep{\fill}}c@{\extracolsep{\fill}}c@{\extracolsep{\fill}}c@{\extracolsep{\fill}}c@{\extracolsep{\fill}}c@{\extracolsep{\fill}}c@{\extracolsep{\fill}}}
\hline
Impactor mass & Impact parameter & Final semi-major axis & Impact \& & Close encounter \& & Ejection  & Triple \\
$M_i$ ($M_e$) & $b$ & $a_f$ & system preservation & system preservation &  & merger \\
\hline
0.01 & 0 & 0.998 & 100 & 0 & 0 & 0 \\ 
0.01 & 0.0556 & 0.995 & 100 & 0 & 0 & 0 \\ 
0.01 & 0.5 & 1 & 0 & 100 & 0 & 0 \\ 
0.01 & 1 & 1 & 0 & 99.7 & 0.3 & 0 \\ 
0.01 & 10 & 1 & 0 & 100 & 0 & 0 \\ 
0.025 & 0 & 1.01 & 100 & 0 & 0 & 0 \\ 
0.025 & 0.0556 & 1.02 & 100 & 0 & 0 & 0 \\ 
0.025 & 0.5 & 1 & 0 & 99.9 & 0.1 & 0 \\ 
0.025 & 1 & 1 & 0 & 99.8 & 0.2 & 0 \\ 
0.025 & 10 & 1 & 0 & 100 & 0 & 0 \\ 
0.05 & 0 & 1.09 & 100 & 0 & 0 & 0 \\ 
0.05 & 0.0556 & 1.12 & 100 & 0 & 0 & 0 \\ 
0.05 & 0.5 & 1.01 & 0 & 99.9 & 0.1 & 0 \\ 
0.05 & 1 & 1 & 0 & 99.8 & 0.2 & 0 \\ 
0.05 & 10 & 1 & 0 & 100 & 0 & 0 \\ 
0.1 & 0 & 3.14 & 88.5 & 0 & 0.1 & 11.4 \\ 
0.1 & 0.0556 & 3.16 & 87.4 & 0 & 0 & 12.6 \\ 
0.1 & 0.5 & 1.02 & 0 & 100 & 0 & 0 \\ 
0.1 & 1 & 1.04 & 0 & 99.4 & 0.6 & 0 \\ 
0.1 & 10 & 1 & 0 & 100 & 0 & 0 \\

\hline
\end{tabular*}
\caption{Probability of various outcomes for an impactor interacting with a proto-Earth$-$moonlet binary system, for an impactor with velocity $v_{imp} = 1.4 v_{esc}$. The mass of the pre-existing satellite is 0.3 $M_l$ and the impactor mass is given by $M_i$ in units of Earth masses. The impact parameter $b$ is in units of the original separation of the proto-Earth--satellite system, which is 18 $R_e$. The outcome of the collision is given as the fractional occurrence of each of four outcomes: an impact followed by preservation of the binary (impact and system preservation), no impact with preservation of the binary (close encounter and system preservation), impact or close encounter followed by ejection of the satellite (ejection), or a triple merger. The average final separation of the proto-Earth--satellite system after the interaction is given by $a_f$, in units of the original separation.}
\label{tbl:fewbody}
\end{table*}

The preservation of the proto-Earth--moonlet systems over a wide range of parameters indicates that impactors of mass $\lesssim$ 0.05 or 0.1 $M_e$ are not large enough to cause significant disruption during collisions or close encounters, respectively, for $v_{imp}=1.4v_{esc}$. We also examined impactors with velocities $v_{imp}=$ 1, 2, and 4 $v_{esc}$. For $v_{imp}=v_{esc}$, most encounters resulted in a collision and preservation of the initial binary, with little change in the semi-major axis or inclination of the pre-existing moonlet. For $v_{imp}$ = 2 (or 4) $v_{esc}$, collisions from impactors of mass $M_i \ge$ 0.05 (or 0.025) were more disruptive, resulting in an increased semi-major axis of the pre-existing moonlet. Faster impacts also resulted in larger changes in the inclination of the pre-existing moonlet, with average changes in inclination of $\lesssim 30^{\circ}$ for all cases except for $0.1 M_e$ mass impactors with $v_{imp} = 4 v_{esc}$, which had an average inclination change of $\sim 73^{\circ}$. However, impacts of such high velocities are likely rare; the average impact velocity of giant impacts in simulations from \citep{Raymond2009a} is $\sim 1.14 v_{esc}$. 

We repeated these tests with a pre-existing moonlet of mass 0.1 and 0.5 $M_l$, and obtained similar results. This implies that while pre-existing moonlets are generally stable during subsequent collisions, impactors $\gtrsim$ 0.1 $M_e$ could disrupt pre-existing satellites during planetary formation, as could impactors with velocities $v_{imp} \ge 2 v_{esc}$, and the timing of when the last impact of mass $M_i \gtrsim$ 0.1 $M_e$ or $v_{imp} \ge 2 v_{esc}$ occurs may coincide with when multiple moonlet systems become stable. A sequence of giant impacts occurring after this could produce a series of new moonlets that form in a system with a stable pre-existing moonlet. Because the pre-exisiting moonlet is likely to migrate to distances $\ge 5 R_e$ (Earth radii) before a subsequent impact, it is unlikely that the pre-existing moonlet will affect the relatively quick formation of a new moonlet from an impact-generated debris disk, based on previous simulations \citep{Citron2014a}. Whether or not the Moon can form from a sequence of such events depends on the likelihood that the pre-existing and newly formed moonlets merge. 

\section{Merger efficiency}

\subsection{Methods}
To estimate the merger probability of moonlets, we modeled a two-moonlet system with one outer moonlet (semi-major axis $a > 10 R_e$) and one inner moonlet ($a = 3.8 R_e$). This approximates a system where a prior moonlet (produced from an earlier impact or moonlet merger) has tidally evolved outward over millions of years before a subsequent giant impact forms an inner moonlet just outside the Roche limit. We examined the dynamical evolution of such two-moonlet systems for a variety of initial conditions to determine moonlet merger probability for systems typical of the multiple-impact hypothesis. 

To evolve the two moonlet system, we used the Mercury-T $N$-body code \citep{Bolmont2015}, which directly integrates the tidal evolution of multi-body systems. Mercury-T accounts for rotation-induced flattening and uses the constant time lag model to compute tidal evolution. We used the proto-Earth as the central body and added the Sun as an external perturber to the system. The Mercury-T code is based on the earlier Mercury $N$-body code \citep{Chambers1999} that treats collisions as perfect mergers. We modified the Mercury-T code so the spin of a body produced by a merger is the sum of the spins of the two bodies that merged. We also treated events where a moonlet's periapse was below the fluid Roche limit of 2.84 $R_e$ as an infall (collision of a moonlet with the proto-Earth). 

For the nominal case (Run 1), we used initial moonlet masses of 0.1 to 0.5 $M_l$ based on typical post-impact debris disk masses \citep{Rufu2017}, and set the outer moonlet's semi-major axis to between 10 and 20 $R_e$ based on the expected orbital migration in the $\sim$ 6 Myrs between large impacts. The inner moonlet was given an initial oribital distance of 3.8 $R_e$, just outside the Roche limit. The eccentricity of both moonlets was set to 0 to account for eccentricity damping and formation on a circular orbit in a disk. The inclination of each moonlet was selected randomly from a uniform distribution of $\cos (i)$ between -1 and 1. Each moonlet was given a random argument of perihelion, longitude of ascending node, and true anomaly. The proto-Earth was given an initial spin axis orientation perpendicular to the orbital plane of the inner moonlet, to reflect that the most recent giant impact can set a planet's spin-orientation \citep{Dones1993b}, and an initial rotation period of 6 hrs \citep{Rufu2017}. 

For Earth, we used $I/MR^2=0.3308$ \citep{Williams1994}, degree-2 potential love number $k_2=0.299$ and fluid love number $k_{2f}=0.993$ \citep{Yoder1995a}, and time lag parameter $k_2 \Delta t=194.59$ s \citep{NerondeSurgy1997}. For moonlets, we used the parameters for the Moon, $I/MR^2=0.394$ \citep{Williams2014}, $k_2=0.0234$ and $k_{2f}=1.440$ \citep{Weber2011,Williams2014}, and $k_2 \Delta t=213$ s \citep{NerondeSurgy1997}.

To examine the merger probability for systems where mergers have already occurred, we conducted additional runs with outer moonlets that are more massive and initially reside at larger semi-major axes (Table \ref{tbl:mercury}).  

\captionsetup{belowskip=-16pt}

\begin{table}
\begin{center}
\begin{tabular*}{\columnwidth}{@{\extracolsep{\fill}}c@{\extracolsep{\fill}}c@{\extracolsep{\fill}}c@{\extracolsep{\fill}}c@{\extracolsep{\fill}}c@{\extracolsep{\fill}}}
\hline
run & $M_{\mathsf{inner}}$ ($M_l$) & $M_{\mathsf{outer}}$ ($M_l$) & $a_{\mathsf{outer}}$ ($R_e)$ & $N_{\mathsf{sims}}$ \\
\hline
1 & 0.1--0.5	& 0.1--0.5	& 10--20 & 490	\\
2 & 0.1--0.3	& 0.3--0.5	& 15--30 & 564	\\
3 & 0.1--0.3	& 0.5--0.8	& 20--30 & 271	\\
\hline
\end{tabular*}
\caption{Parameters used in Mercury-T simulations.}
\label{tbl:mercury}
\end{center}
\end{table}
\captionsetup{belowskip=+6pt}

\subsection{Results}

\begin{figure}[h] 
\centerline{\includegraphics[trim={0 0.5cm 0 1.3cm},clip,scale=0.38]{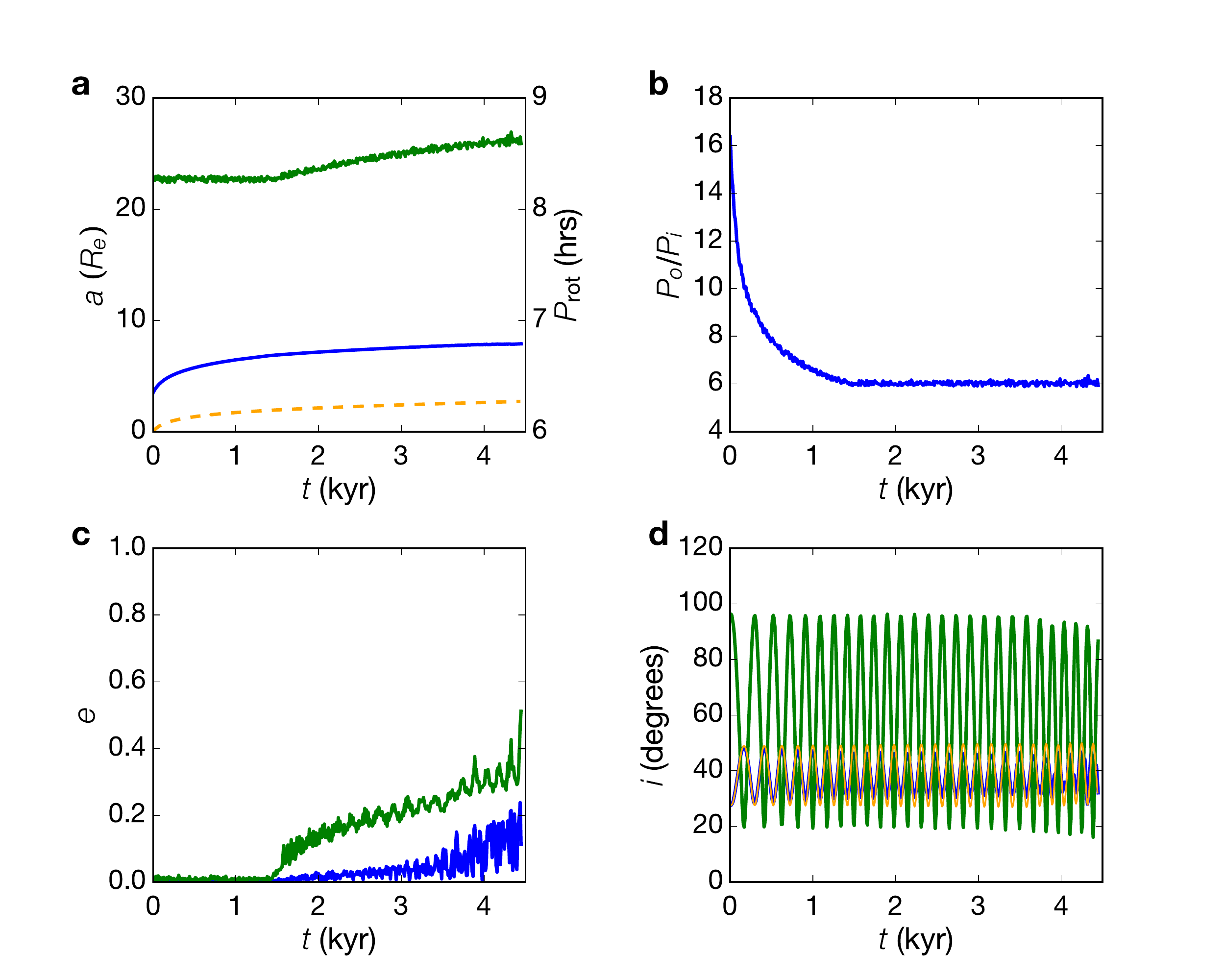}}
\caption{Example of a typical simulation that results in a merger. Evolution over time is shown for (a) semi-major axis, (b) period ratio, (c) eccentricity, and (d) inclination. The initial mass of the inner (blue) and outer (green) moonlets is 0.26 and 0.32 $M_l$ (lunar masses), respectively. The rotational period and obliquity of the proto-Earth is shown in orange in subplots (a) and (d). Initially, the inner moonlet migrates outward at a faster rate than the outer moonlet, as expected. However, the inner moonlet is relatively quickly captured into a 6:1 resonance with the outer moonlet. This increases the eccentricity and migration rate of the outer moonlet. Despite the outer moonlet's faster migration rate relative to the inner moonlet, the increased eccentricity of the outer moonlet leads to a collision at $\sim$ 4.5 kyr. After the merger, the resulting 0.58 $M_l$ moonlet tidally evolves outward (not shown).} 
\label{fig:sim1}
\end{figure}

\begin{figure}
\centerline{\includegraphics[trim={0 0.5cm 0 1.5cm},clip,scale=0.38]{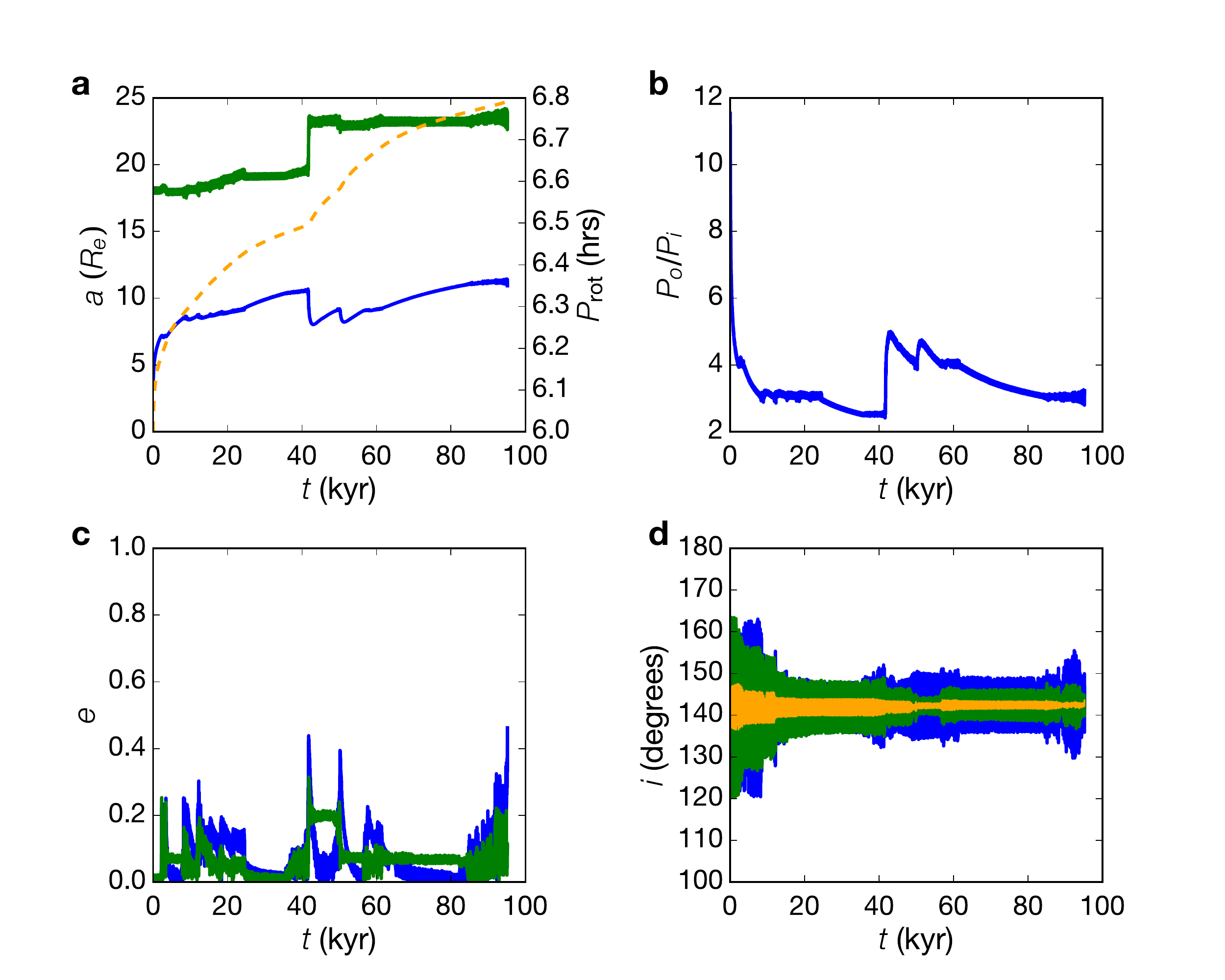}}
\caption{An additional example of a simulation that resulted in a merger. Evolution over time is shown for (a) semi-major axis, (b) period ratio, (c) eccentricity, and (d) inclination. The initial mass of the inner (blue) and outer (green) moonlets is 0.23 and 0.36 $M_l$, respectively. The rotational period and obliquity of the proto-Earth is shown in orange in subplots (a) and (d). In this simulation, the system passes through multiple resonances that result in close encounters, before a final merger at $\sim$ 95 Myr.} 
\label{fig:sim2}
\end{figure}

\begin{figure}
\centerline{\includegraphics[trim={0 0.5cm 0 1.5cm},clip,scale=0.38]{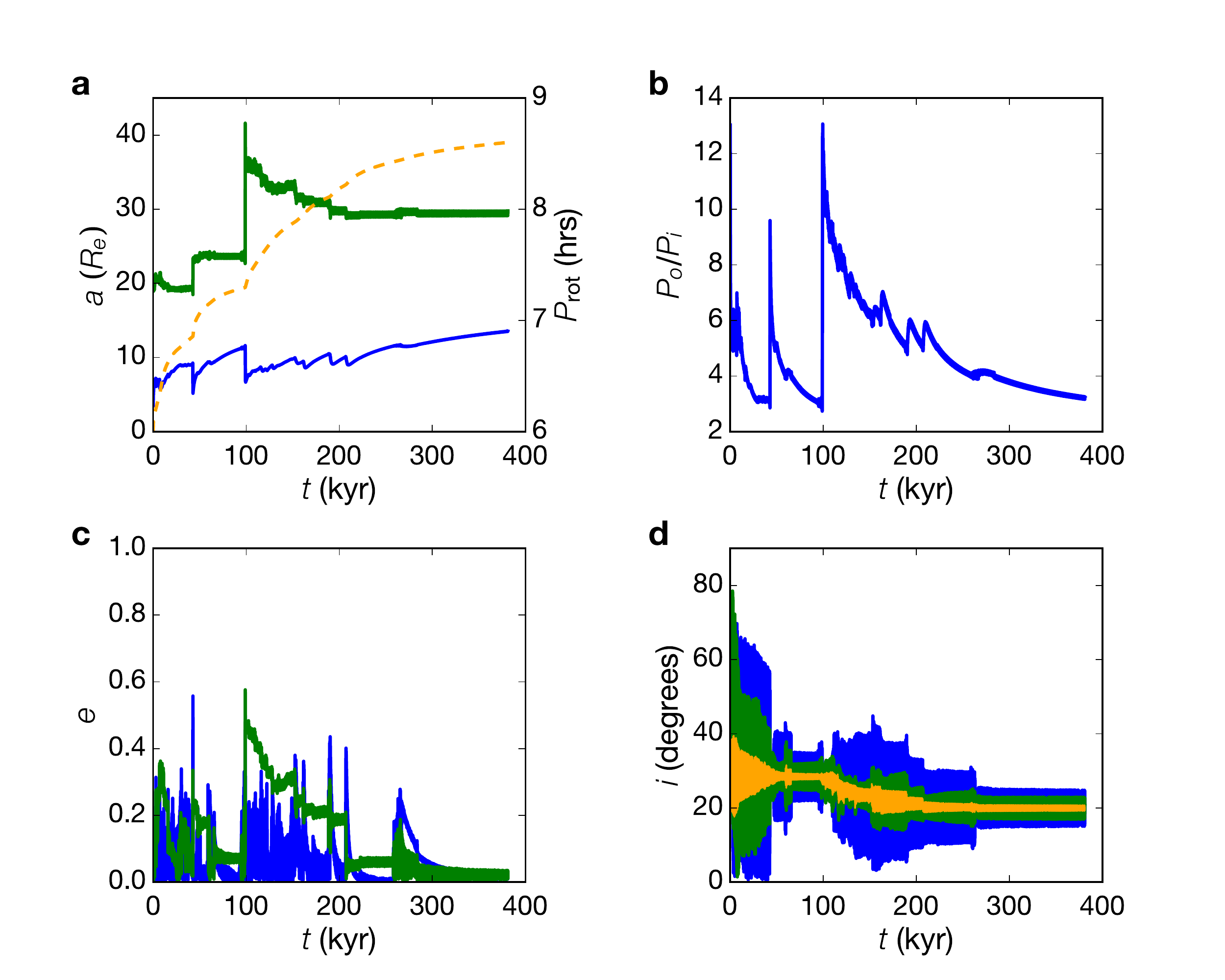}}
\caption{Example output for a simulation that resulted in both moonlets infalling (colliding with the proto-Earth). The initial mass of the inner (blue) and outer (green) moonlets is 0.23 and 0.32 $M_l$, respectively. The rotational period and obliquity of the proto-Earth is shown in orange in subplots (a) and (d). The system evolves through several resonances and close encounters before a close encounter results in both moonlets infalling.} 
\label{fig:sim3}
\end{figure}

We find that mergers between two moonlets are possible, but the preceding orbital evolution can be quite complex. Similar to Fig. \ref{fig:sim1}, inner moonlets evolved outward until they were captured into a resonance, generally between 2:1 to 6:1. These resonances increased the eccentricity of the orbits, resulting in close encounters and either an eventual merger or system disruption. Often, the system passed through multiple resonances before a disruptive close encounter or merger (Figs. \ref{fig:sim2} and \ref{fig:sim3}), suggesting that analytical methods of computing tidal evolution may be insufficient for assessing moonlet mergers and dynamics. The orbital evolution of moonlets is also affected by secular processes such as Lidov-Kozai \citep{Kozai1962} evolution (in the Sun-Earth-moonlet triple), and its coupling to the effects of the planet obliquity and tidal evolution.

\begin{figure}
\centerline{\includegraphics[trim={0 0.5cm 0 0cm},clip,scale=0.65]{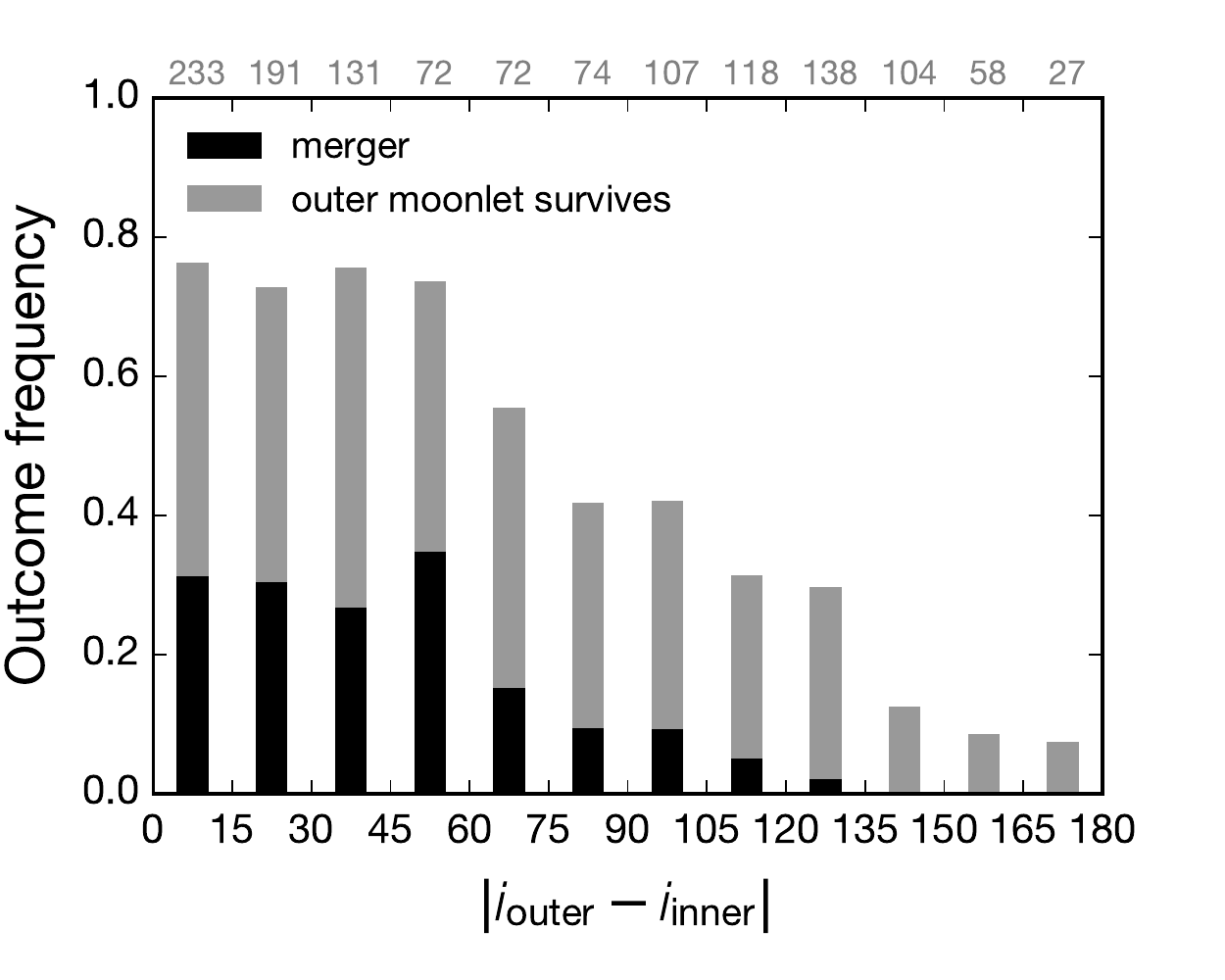}}
\caption{Histogram showing the merger probability (solid bars) and outer moonlet survival probability (stacked shaded bars) versus the difference in inclination between the outer and inner moonlet. The cumulative result for all three runs is plotted. Simulations per bin are indicated above each bar. } 
\label{fig:full_idiff}
\end{figure}

\begin{figure} 
\centerline{\includegraphics[trim={0 1.5cm 0 0cm},clip,scale=0.33]{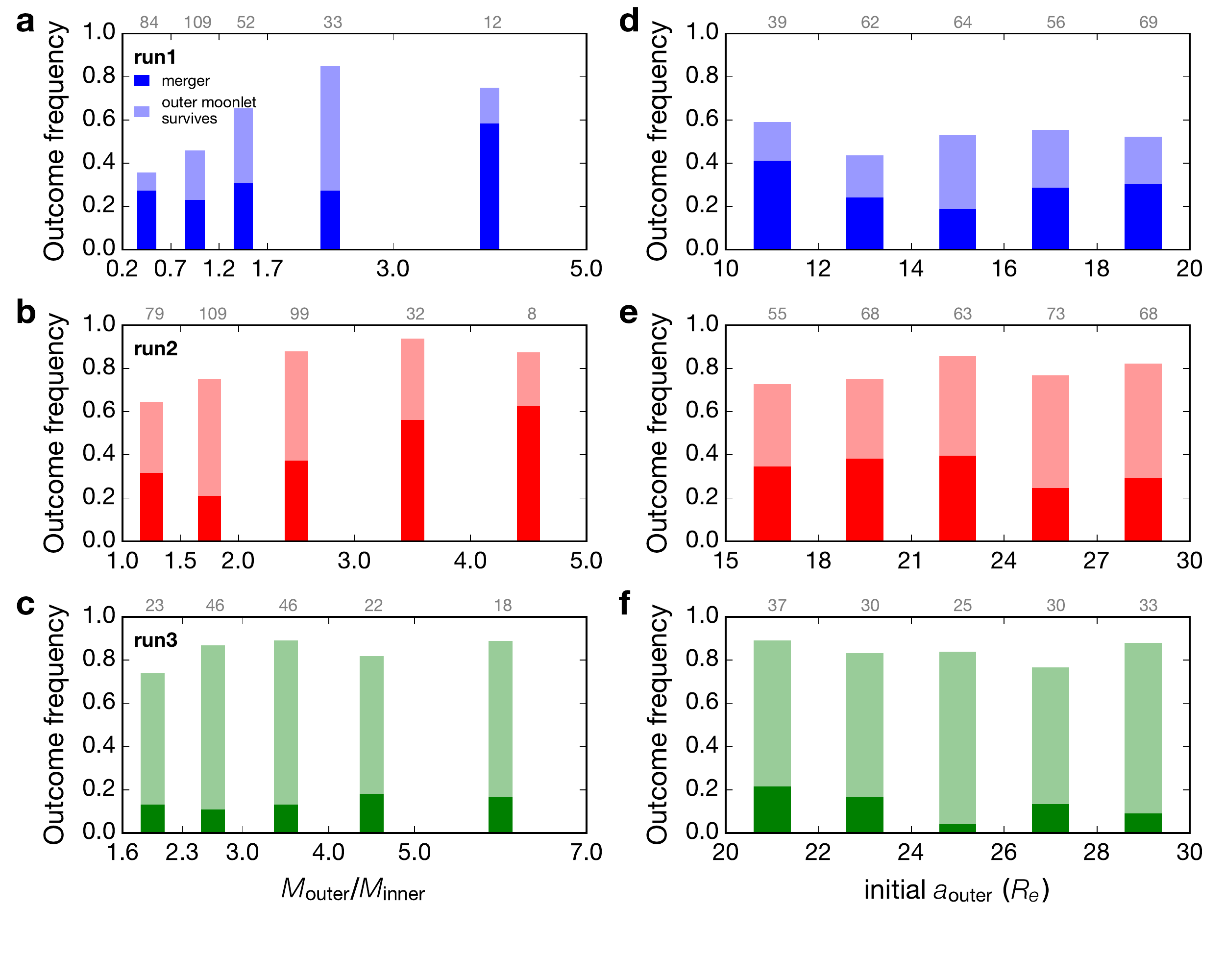}}
\caption{Histograms showing the merger probability versus mass ratio (a-c) and initial outer moon semi-major axis (d-f), for prograde-prograde simulations. Solid bars indicate the merger probability and stacked shaded bars indicate system survival probability (merger or outer moonlet survival). Simulations per bin are indicated above each bar. Results for runs 1, 2, and 3, are plotted in the top, middle, and bottom rows, respectively. }
\label{fig:prograde_mass_a2}
\end{figure}

To evaluate the feasibility of the multiple impact hypothesis, we examined the likelihood that the inner and outer moonlets merged into a stable larger moonlet (the merger probability), and the likelihood that the outer moonlet remained stable even if the inner moonlet was ejected or impacted the proto-Earth (the outer moonlet survival probability). The merger probabilities for the simulations outlined in Table \ref{tbl:mercury} depend non-trivially on the initial configuration of the moonlet systems. Systems containing both a retrograde moonlet and a prograde one, for example, were far more prone to be destroyed (see Fig. \ref{fig:full_idiff}; mostly through collisions of the moonlets with the Earth). However, it is difficult to produce subsequent moonlets with a different sense of rotation than the pre-existing moonlet \citep{Rufu2017}. Therefore, we expect both moonlets to have same-sense orbits. For systems in which both moonlets were prograde, the probability of outer moonlet survival increases with outer moonlet mass and initial semi-major axis (Fig. \ref{fig:prograde_mass_a2}). The probability of merger generally increases with outer moonlet mass (Fig. \ref{fig:prograde_mass_a2}a,b), but decreases if the outer moonlet was too massive and distant (Fig. \ref{fig:prograde_mass_a2}c,f). 

\begin{figure}
\centerline{\includegraphics[trim={0 3cm 0 0cm},clip,scale=0.31]{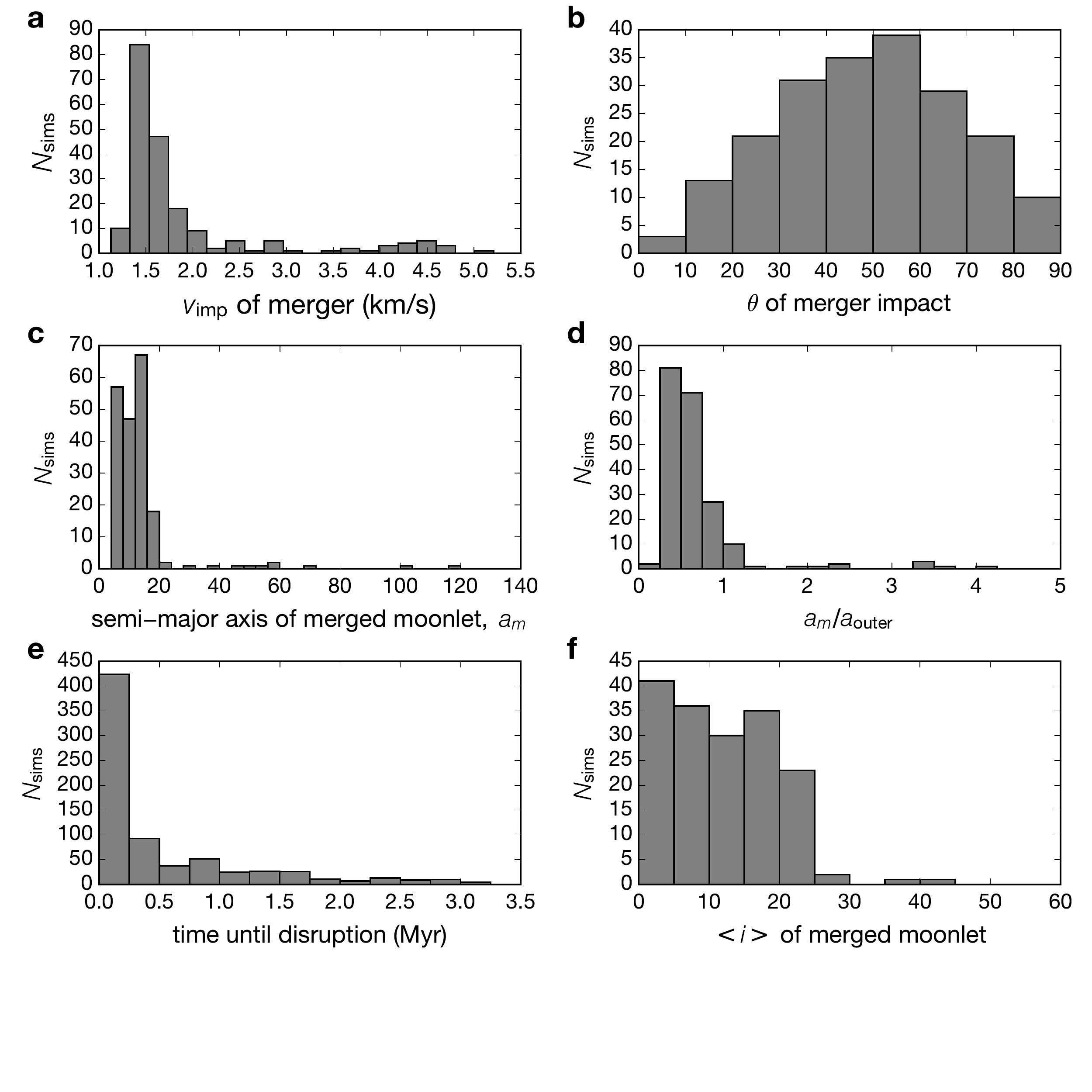}}
\caption{Histograms of (a) impact velocity, (b) impact angle of merging impacts between moonlets, (c) the initial semi-major axis of the product of a merger between two moonlets, $a_m$, computed immidiately after the merger, (d) the ratio of $a_m$ to the initial semi-major axis of the outer moonlet, $a_{\mathsf{outer}}$, (e) the time a two-moonlet system evolves until a moonlet is destroyed or the moonlets merge, and (f) the mean inclination of each merged moonlet after the merger, for simulations in which both moonlets were prograde (the mean standard deviation of the post-merger inclination was $\sim 6^{\circ}$.}
\label{fig:combined}
\end{figure}

For stable mergers, the average merging impact velocity was $\sim$ 1.9 km/s (Fig. \ref{fig:combined}a,b). Non-merger simulations were disrupted by close encounters between the two moonlets or a moonlet and the proto-Earth, most commonly resulting in one or both of the moonlets de-orbiting and impacting the proto-Earth (sometimes after merging). Only rare cases resulted in moonlet ejection. 

From a dynamical perspective, the feasibility of the multiple-impact hypothesis for Moon formation depends on the likelihood of continued moonlet growth. While the merger probability is low for large, distant moonlets (semi-major axis $>$20 $R_e$), moons produced from mergers typically had closer orbits (semi-major axis $<$20 $R_e$), less than the original outer moonlet (Fig. \ref{fig:combined}c,d). Disruptions generally occurred within the first 500 kyr of the simulation (Fig. \ref{fig:combined}e), making it unlikely that two moonlets could coexist until a subsequent impact (in $\sim$ 6 Myr). While less massive outer moonlets could inhibit system growth (Fig. \ref{fig:prograde_mass_a2}a), if the outer moonlet is at least twice as massive as the inner moonlet, $1-4$ mergers are expected to occur before system disruption (Fig. \ref{fig:prograde_mass_a2}a,b), for moderate mass outer moonlets. Such a small number of mergers is less than the 20 impacts suggested by the Monte Carlo simulations of \citet{Rufu2017}. However, a sequence of such mergers could still aid in producing a more compositionally similar Earth and Moon if the impactors are sufficiently small and fast to eject a greater fraction of target material \citep{Rufu2017}. For more massive outer moonlets (Run 3; Fig. \ref{fig:prograde_mass_a2}c), moonlet mergers become less likely, implying that growth of the outer, pre-exisiting moonlet through mergers could inhibit later growth. However, in this scenario the probability of outer moonlet survival remains high, implying that for large pre-existing moons, several giant impacts could post-date moon formation.

\captionsetup{belowskip=+42pt}

\captionsetup{belowskip=-42pt}

\begin{figure}
\centerline{\includegraphics[trim={0 1.5cm 0 0cm},clip,scale=0.33]{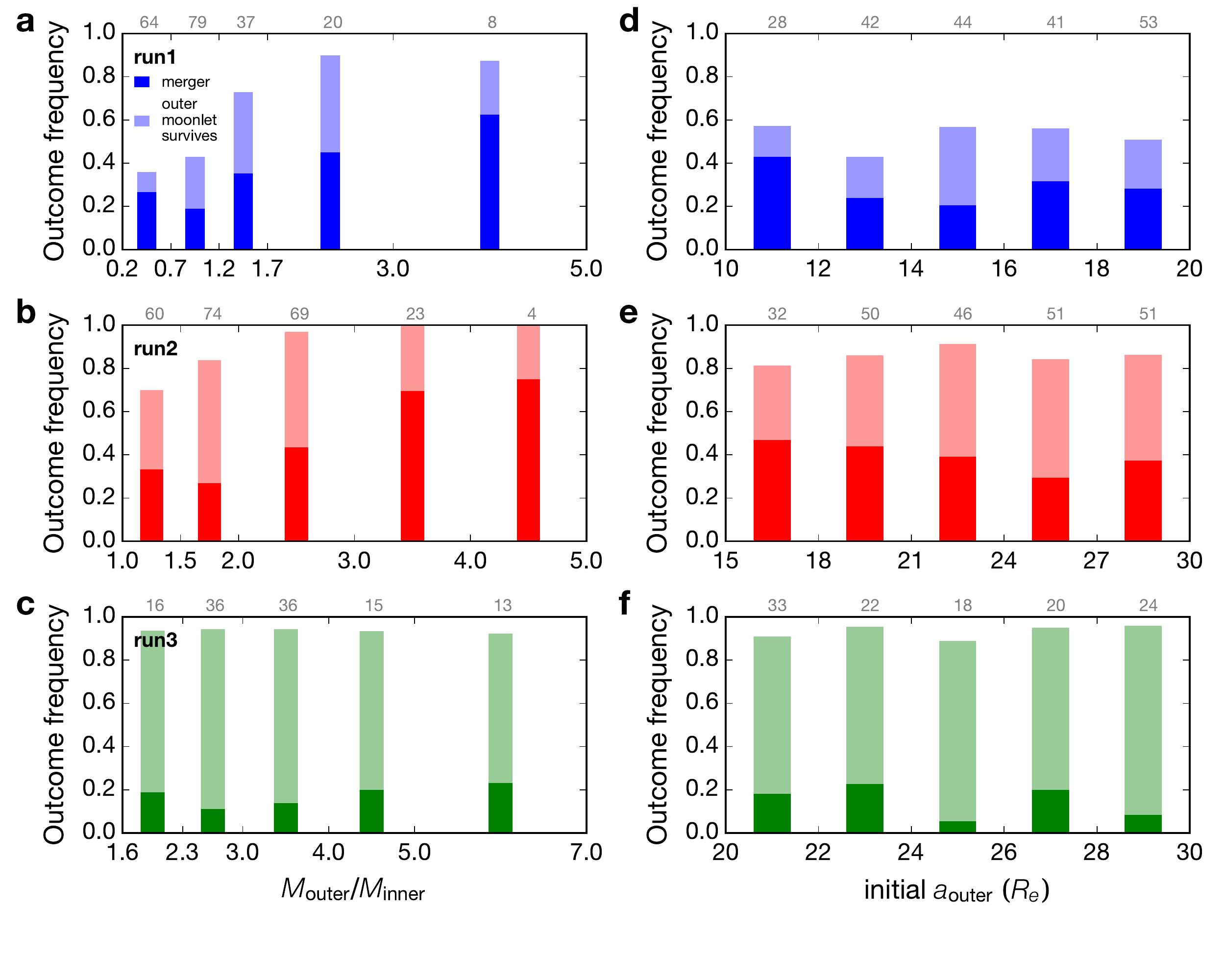}}
\caption{Histograms showing the merger probability versus mass ratio (a-c) and initial outer moon semi-major axis (d-f), for prograde-prograde simulations where $|i_{\mathsf{outer}}-i{_\mathsf{inner}}|<45^{\circ}$. Solid bars indicate the merger probability and stacked shaded bars indicate system survival probability (merger or outer moonlet survival). Simulations per bin are indicated above each bar. Results for runs 1, 2, and 3, are plotted in the top, middle, and bottom rows, respectively. }
\label{fig:prograde_45_mass_a2}
\end{figure}

\clearpage

\section{Discussion}

The merger probability we estimate for prograde-prograde systems may in a certain sense be considered a lower limit. For simplicity, we set the obliquity of the proto-Earth to match the inclination of the inner moonlet, the same as the last impact. This is not necessarily the case for a sufficiently spun-up proto-Earth and smaller impactors, and even retrograde impacts can produce prograde moonlets (if any) \citep{Rufu2017}. If the Laplace plane of the pre-existing moonlet is close to equator of a fast-rotating proto-Earth, then the inclination of a small moonlet produced from a subsequent impact would likely be closer to that of the pre-existing moonlet than a random distribution of inclinations would suggest. This would increase the chance of building a Moon-sized body from multiple impacts, because the merger probability and system survival probability are higher when the moonlets have more similar inclinations (Figs. \ref{fig:full_idiff} and \ref{fig:prograde_45_mass_a2}). Additionally, the merger probability would slightly increase if moonlets are allowed to achieve a pericenter interior of the Roche limit, instead of being disrupted at the Roche limit of 2.84 $R_e$ as we assume; based on \citet{Sridhar1992}, a moonlet might survive passage through a periapsis as close as $\sim$ 2 $R_e$. We also neglect the possibility that tidally disrupted moonlets might also form a disk that generates a new moonlet (or several generations of new moonlets, {\it e.g.}, \citet{Hesselbrock2017}), which could eventually merge with the non-disrupted moonlet and increase the overall merger probability.

The parameter space associated with giant impacts and moonlet formation during the accretion of the Earth is immense, and our study was limited in the number of impact scenarios and initial moonlet configurations that we could explore. Results could be affected by changing the starting position of the outer moonlet, which depends on the time between large impacts. Merger probability could also be sensitive to the tidal dissipation of the proto-Earth and moonlets. \citet{Zahnle2015} suggest the Earth would be weakly dissipative for several Myr following a giant impact; while the impacts we consider are not as energetic, a weakly dissipative proto-Earth would delay the time until the moonlets reach a mutual resonance and subsequently merge or infall. We also assumed a fixed proto-Earth mass, which would actually begin smaller and increase as impacts add material and more moonlets are formed. 

Our examination of the tidal evolution of two-moonlet systems suggests that the Moon could be a product of one or more moonlet mergers. For prograde-prograde systems the probability of a merger or outer moonlet survival is $\sim$ 70\%, and generally increases when the outer moonlet is larger. Additionally, the low velocity at which the moonlets merge ($\sim$ 1.9 km/s) could explain some of the heterogeneities in the Moon's interior ({\it e.g.}, \citet{Robinson2016}). \citet{Jutzi2011} examined the generation of the lunar highlands from a similar low velocity (2$-$3 km/s) impact, but in the context of two moons produced concurrently from a single impact. However, the low velocity collision they discuss could also occur from the last moonlet merger in a multiple-impact Moon formation scenario. We also find that the average inclination of merged moonlets following the merger was 10.7$^{\circ}$, with $\sim$ 50\% of mergers resulting in average inclinations $>$9$^{\circ}$ (Fig. \ref{fig:combined}f). Moonlet mergers could therefore provide a solution to the mutual inclination problem, which suggests that the Moon began with an inclination of $\sim$12$^{\circ}$ \citep{Touma1998}. 

While we conducted our simulations to examine the multiple-impact hypothesis, our results also have implications for Moon formation in a single-impact scenario. Our findings indicate that the early Earth-Moon system could have been stable against a subsequent large impactor of mass $M_i \leq 0.05 M_e$ (Earth masses), because proto-Earth$-$moonlet binaries are preserved during such collisions, and disruptions of large and distant moonlets via subsequent orbital interactions with smaller inner moonlets are rare (Fig. \ref{fig:prograde_mass_a2}). This implies that while a single Moon-forming impact could have been the largest late impact to occur, it may have been followed by other large collisions. Examination of the likelihood of late Moon-forming impacts ({\it e.g.}, \citet{Jacobson2014a}), should consider the last several giant impacts, which could increase the likelihood of single Moon-forming collisions. Additionally, since large impacts remove angular momentum from the system \citep{Rufu2017}, collisions occurring after a single Moon-forming impact could allow for an Earth-Moon system with a higher initial angular momentum, as invoked in several Moon-formation impact scenarios ({\it e.g.}, \citet{Cuk2012}, \citet{Cuk2016}, \citet{Canup:2012}, and \citet{Reufer2012}). Our simulations might also explain why some large terrestrial planets, such as Venus, might not have satellites, because pre-existing satellites often de-orbit after merging with (or in addition to) a newly generated moonlet. This could also explain the lack of significant moons around terrestrial planets in extra-solar planetary systems.

\clearpage
\section{Conclusions}

The abundance of giant impacts (each capable of producing a debris disks and satellite) in the late stages of planetary formation suggests that multiple-satellite systems may have been a common occurrence. Our simulations show that the Earth may have had several past moons. If prior moonlets merged, the Moon could have formed from a sequence of giant impacts, which could explain the similarity in isotopic composition to the Earth. And in the context of typical giant impacts during planetary formation, it may be more likely for several small, fast impactors to eject sufficient material into proto-Earth orbit to form the Moon than a single impact with finely-tuned parameters. Sequences of impacts that result in 1$-$4 moonlet mergers are possible, particularly if the outer moonlet is larger than the inner moonlet and at an intermediate distance from the proto-Earth, and the likelihood of moonlet mergers could increase if subsequent moonlets are preferentially generated near the Laplace plane of pre-existing moonlets. Our simulations suggest that moonlets are also likely to infall and impact the proto-Earth, which could have consequences for early Earth evolution \citep{Malamud2018}. In the context of a single-impact origin of the Moon, we find that several giant impacts may have post-dated Moon formation, because a large outer moon is stable against subsequent impacts and moonlet formation events. Subsequent impacts could therefore provide another means of removing angular momentum from the Earth-Moon system, allowing impact scenarios that rely on a higher post-impact angular momentum state.

\section*{ }

The authors wish to thank the Minerva Center for Life Under Extreme Planetary Conditions, the I-CORE Program of the PBC and ISF (Center No. 1829/12) and the Helen Kimmel Center for Planetary Sciences for support of this work. O.A. wishes to thank the Adolf and Mary Mil Stiftung Foundation and the IMOS (Grant No. 712626) for their support. We also thank Raluca Rufu and Tushar Mittal for helpful discussions. We thank one anonymous reviewer for comments and suggestions that improved the manuscript.

\bibliographystyle{aasjournal}
\bibliography{moon_paper_bib}

\appendix

\section{Supplementary Tables}

\begin{table*}[h!]
\begin{center}
\begin{tabular*}{\columnwidth}{@{\extracolsep{\fill}}c@{\extracolsep{\fill}}c@{\extracolsep{\fill}}c@{\extracolsep{\fill}}c@{\extracolsep{\fill}}c@{\extracolsep{\fill}}c@{\extracolsep{\fill}}c@{\extracolsep{\fill}}}
\hline
Impactor mass & Impact parameter & Final semi-major axis & Impact \& & Close encounter \& & Ejection  & Triple \\
$M_i$ ($M_e$) & $b$ & $a_f$ & system preservation & system preservation &  & merger \\
\hline
0.01 & 0 & 1 & 100 & 0 & 0 & 0 \\ 
0.01 & 0.0556 & 1.01 & 100 & 0 & 0 & 0 \\ 
0.01 & 0.5 & 1 & 0 & 100 & 0 & 0 \\ 
0.01 & 1 & 1 & 0 & 100 & 0 & 0 \\ 
0.01 & 10 & 1 & 0 & 100 & 0 & 0 \\ 
0.025 & 0 & 1.12 & 99.9 & 0 & 0 & 0.1 \\ 
0.025 & 0.0556 & 1.12 & 100 & 0 & 0 & 0 \\ 
0.025 & 0.5 & 1 & 0 & 99.9 & 0.1 & 0 \\ 
0.025 & 1 & 1 & 0 & 99.8 & 0.2 & 0 \\ 
0.025 & 10 & 1 & 0 & 100 & 0 & 0 \\ 
0.05 & 0 & 2.53 & 93.7 & 0 & 0 & 6.3 \\ 
0.05 & 0.0556 & 2.46 & 93.2 & 0 & 0 & 6.8 \\ 
0.05 & 0.5 & 1 & 0 & 100 & 0 & 0 \\ 
0.05 & 1 & 1 & 0 & 99.6 & 0.4 & 0 \\ 
0.05 & 10 & 1 & 0 & 100 & 0 & 0 \\ 
0.1 & 0 & 3.79 & 42 & 0 & 35.6 & 22.4 \\ 
0.1 & 0.0556 & 5.96 & 45.4 & 0 & 34.4 & 20.2 \\ 
0.1 & 0.5 & 1.02 & 0 & 99.9 & 0.1 & 0 \\ 
0.1 & 1 & 1.01 & 0 & 99.6 & 0.4 & 0 \\ 
0.1 & 10 & 1 & 0 & 100 & 0 & 0 \\ 

\hline
\end{tabular*}
\caption{As in Table 1, but for $v_{imp} = 2 v_{esc}$.}
\label{tbl:fewbody2}
\end{center}
\end{table*}


\clearpage

\begin{table*}[h!]
\begin{center}
\begin{tabular*}{\columnwidth}{@{\extracolsep{\fill}}c@{\extracolsep{\fill}}c@{\extracolsep{\fill}}c@{\extracolsep{\fill}}c@{\extracolsep{\fill}}c@{\extracolsep{\fill}}c@{\extracolsep{\fill}}c@{\extracolsep{\fill}}}
\hline
Impactor mass & Impact parameter & Final semi-major axis & Impact \& & Close encounter \& & Ejection  & Triple \\
$M_i$ ($M_e$) & $b$ & $a_f$ & system preservation & system preservation &  & merger \\
\hline
0.01 & 0 & 1.12 & 100 & 0 & 0 & 0 \\ 
0.01 & 0.0556 & 1.09 & 100 & 0 & 0 & 0 \\ 
0.01 & 0.5 & 1 & 0 & 99.9 & 0.1 & 0 \\ 
0.01 & 1 & 1 & 0 & 100 & 0 & 0 \\ 
0.01 & 10 & 1 & 0 & 100 & 0 & 0 \\ 
0.025 & 0 & 5.63 & 81.9 & 0 & 9.4 & 8.7 \\ 
0.025 & 0.0556 & 4.36 & 82.6 & 0 & 9.1 & 8.3 \\ 
0.025 & 0.5 & 1 & 0 & 100 & 0 & 0 \\ 
0.025 & 1 & 1 & 0 & 100 & 0 & 0 \\ 
0.025 & 10 & 1 & 0 & 100 & 0 & 0 \\ 
0.05 & 0 & 5.28 & 32.9 & 0 & 46.7 & 20.4 \\ 
0.05 & 0.0556 & 4.75 & 33.9 & 0 & 46.8 & 19.3 \\ 
0.05 & 0.5 & 1 & 0 & 100 & 0 & 0 \\ 
0.05 & 1 & 1 & 0 & 99.6 & 0.4 & 0 \\ 
0.05 & 10 & 1 & 0 & 100 & 0 & 0 \\ 
0.1 & 0 & 5.1 & 15.4 & 0 & 76.2 & 8.4 \\ 
0.1 & 0.0556 & 4.24 & 14.9 & 0 & 76 & 9.1 \\ 
0.1 & 0.5 & 1 & 0 & 99.9 & 0.1 & 0 \\ 
0.1 & 1 & 1 & 0 & 99.9 & 0.1 & 0 \\ 
0.1 & 10 & 1 & 0 & 100 & 0 & 0 \\ 

\hline
\end{tabular*}
\caption{As in Table 1, but for $v_{imp} = 4 v_{esc}$.}
\label{tbl:fewbody3}
\end{center}
\end{table*}

\end{document}